\documentclass[%
reprint,
superscriptaddress,
showpacs,preprintnumbers,
nobibnotes,
amsmath,amssymb,
aps,
prl,
notitlepage
]{revtex4-1}

\usepackage{graphicx}
\usepackage[export]{adjustbox}
\usepackage{dcolumn}
\usepackage{bm}
\usepackage{wasysym}
\usepackage[utf8]{inputenc}
\usepackage{color}

\begin{document}
\newlength\fheight \newlength\fwidth 

\title{Scaling behavior of spin transport in hydrogenated graphene}
\author{M. R. Thomsen}
\affiliation{Department of Physics and Nanotechnology, Aalborg University, DK-9220 Aalborg \O st, Denmark\\Center for Nanostructured Graphene (CNG), DK-9220 Aalborg \O st, Denmark}
\author{M. Ervasti}
\affiliation{COMP Centre of Excellence, Department of Applied Physics, Aalto University, Helsinki, Finland}
\author{A. Harju}
\affiliation{COMP Centre of Excellence, Department of Applied Physics, Aalto University, Helsinki, Finland}
\author{T. G. Pedersen}
\affiliation{Department of Physics and Nanotechnology, Aalborg University, DK-9220 Aalborg \O st, Denmark\\Center for Nanostructured Graphene (CNG), DK-9220 Aalborg \O st, Denmark}

\begin{abstract}
We calculate the spin transport of hydrogenated graphene using the Landauer-Büttiker formalism with a spin-dependent tight-binding Hamiltonian. The advantages of using this method is that it simultaneously gives information on sheet resistance and localization length as well as spin relaxation length. Furthermore, the Landauer-Büttiker formula can be computed very efficiently using the recursive Green's function technique. 
Previous theoretical results on spin relaxation time in hydrogenated graphene have not been in agreement with experiments. Here, we study magnetic defects in graphene with randomly aligned magnetic moments, where interference between spin-channels is explicitly included. We show that the spin relaxation length and sheet resistance scale nearly linearly with the impurity concentration. Moreover, the spin relaxation mechanism in hydrogenated graphene is Markovian only near the charge neutrality point or in the highly dilute impurity limit. 
\end{abstract}

\maketitle

\section{Introduction}
Spin transport in graphene has attracted a lot of attention in recent years due to very long spin relaxation times and spin relaxation lengths predicted for this material~\cite{huertas2009spin,han2014graphene}. The spin relaxation length in graphene has been predicted theoretically to be at least 20 $\mu m$~\cite{huertas2009spin}, whereas experimental values are about an order of magnitude lower, typically around 1--4 $\mu m$~\cite{wojtaszek2013enhancement,tombros2008anisotropic,tombros2007electronic,han2011spin,wojtaszek2014absence,zomer2012long}. It has been ruled out experimentally that this discrepancy is due to hyperfine interaction with naturally occurring $^{13}$C isotope in graphene~\cite{wojtaszek2014absence}. An attempt to explain the discrepancy between theory and experiment based on magnetic impurities in graphene has been given by Kochan \emph{et al.}~\cite{kochan2014spin}. Magnetic impurities are very common in graphene and may for instance be hydrogen adatoms~\cite{yazyev2007defect}, vacancies~\cite{ma2004magnetic,yazyev2007defect} or embedded metal atoms~\cite{krasheninnikov2009embedding,thomsen2015stability} in graphene pores. Kochan \emph{et al.} find that 0.36 ppm coverage of hydrogen adatoms is sufficient to obtain spin relaxation times that are in agreement with experiments. Their model is based on the Green's function of a single hydrogen adatom in an infinite graphene sheet and multiplying their results with the impurity concentration. In effect, they do not include interference effects between scatterers in their model and their model is thus only valid in the highly dilute limit. Experimental measurements of graphene in the presence of a strong magnetic field confirms that the observed low spin relaxation length is, at least in part, due to magnetic impurities in graphene~\cite{lara-avila2015influence}.
Spin transport in hydrogenated graphene was also considered by Soriano \emph{et al.}~\cite{soriano2011magnetoresistance,soriano2015spin}. Their method is based on a mean-field Hubbard Hamiltonian and the real space Kubo-transport formalism. They find that a coverage of 15~ppm hydrogen adatoms gives the correct order of magnitude of the spin relaxation time~\cite{soriano2015spin}, which is more than an order of magnitude larger than the prediction by Kochan \emph{et al}. Additionally, the functional form of the two theoretical results does not resemble the experimental result. A recent \textit{ab initio} study of the spin scattering of hydrogen adatoms on narrow armchair graphene nanoribbons by Wilhelm \textit{et al.}~\cite{wilhelm2015ab} has shown that spin scattering off a single hydrogen adatom with defect spin oriented perpendicular to the electron spin is sufficient to obtain spin-flip conductance on the same order of magnitude as the spin-conserved conductance. They also showed spin-orbit interactions to be negligible compared to exchange interactions in the context of spin scattering on hydrogen adatoms.

The spin relaxation length is determined by the decay rate of spin polarization. Zurek \emph{et al.}~\cite{zurek2007gaussian} have found through a phenomenological spin interaction Hamiltonian that the spin relaxation decay rate depends on the distribution of coupling strengths between a spin system and an environment with many independent spins. In particular, they find that a Gaussian distribution of couplings leads to Gaussian decay of the spin polarization with respect to time, whereas a Lorentzian distribution leads to exponential decay. It is straightforward to demonstrate that the spin relaxation of scatterers on a classical Markovian chain is also exponential. Therefore, exponential decay of spin polarization is typically referred to as Markovian behavior \cite{coish2008exponential}.

In this paper, we calculate the spin-dependent electron transport on graphene with hydrogen adatoms using the Landauer-Büttiker formalism, which is a widely used method for calculating quantum transport in nanoscale devices \cite{power2014electronic,thomsen2014dirac,datta1995electronic,wilhelm2015ab,markussen2006electronic,pedersen2012transport}. We use hydrogen adatoms as they are very common magnetic defects on graphene. They have a local magnetic moment of approximately 1 $\mu_B$ per adatom. Additionally, due to local $sp^3$ hybridization, hydrogenated graphene has an energy gap~\cite{balog2010bandgap}. In particular, we will demonstrate that the Landauer-Büttiker formalism can be used to extract the spin relaxation length of a system. We will demonstrate that the spin relaxation is not always Markovian and that spin relaxation length and sheet resistance scales nearly linearly with impurity concentration.

\section{Theoretical Methods}

We employ a third-nearest neighbor $\pi$-electron tight-binding (TB) model to set up a spin-dependent system Hamiltonian on the form 

\begin{equation}
\hat H = \sum_{\langle i,j \rangle,\sigma} t_{ij,\sigma}(\hat c_{i,\sigma}^\dagger \hat c_{j,\sigma}+\text{H.c}) + \sum_{i,\sigma}\varepsilon_{i,\sigma}\hat c_{i,\sigma}^\dagger \hat c_{i,\sigma}\,,
\end{equation}

\noindent where $\hat c_{i,\sigma}^\dagger$ ($\hat c_{j,\sigma}$) is the creation (annihilation) operator on the lattice site $i$ ($j$) with spin $\sigma$, $t_{ij,\sigma}$ is the spin-dependent hopping parameter between site $i$ and $j$ and $\varepsilon_i$ is the spin-dependent on-site energy on site $i$. The notation $\langle i,j \rangle$ is used to represent up to third nearest neighbor indices. 
TB parameters of hydrogen adatoms on graphene are obtained by fitting the TB band structure and local density of states (LDOS) to match the DFT band structure and atom-projected partial DOS (pDOS), respectively. 

The DFT calculations are carried out using the FHI-aims package~\cite{blum2009ab}. It is an all-electron code with numerical atom-centered basis functions. We use the default \textit{tight} basis set for each atom type in a spin-polarized calculation. The electron-electron interactions are treated at the level of the Perdew-Burke-Ernzerhof (PBE) exchange-correlation functional~\cite{perdew1996generalized}. The hydrogen adatom on graphene is relaxed in a supercell with $2 \times 8 \times 8 = 128$ carbon atoms, until the forces between atoms are smaller than $10^{-3}$ eV/\AA. We expect this supercell to be large enough for finite size effects to be negligible. Moreover, the DFT self-consistency cycle is considered converged if, among other things, the total energy changes by less than $10^{-6}$ eV. We use an $8 \times 8 \times 1$ $k$-point Monkhorst-Pack grid under relaxation. The final band structure and pDOS calculations are computed using k-grids of $15 \times 15 \times 1$ and $12 \times 12 \times 1$ $k$-points, respectively. For the band structure fit, we compare the six lowest unoccupied and six highest occupied bands 
and fit the two-dimensional band energies in the first Brillouin zone. 

\begin{figure}
	\includegraphics[width=0.9\columnwidth]{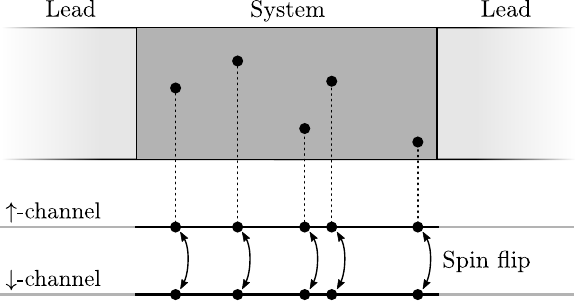}
	\caption{Spin-dependent transport is equivalent to having two separate channels that couple only at magnetic impurity sites.}
	\label{fig:spin_transport_device} 
\end{figure}

The defect spin moment may not be aligned with the chosen quantization axis of the electron spin in the leads. In order to take this into account in our model, we rotate the defect spins in spin space. 
To do so, we start by writing the spin-dependent Hamiltonian as a sum of carbon $\hat H_C$ and defect $\hat H_{\text{defect}}$ parts, $\hat H = \sum_{\text{graphene}} \hat H_{C} + \sum_{\text{defects},i} \hat H_{\text{defect}}^{(i)}$. The defect Hamiltonian is separated into spin channels and is written as $\hat H_{\text{defect},\uparrow/\downarrow}^{(i)}=\bar H_i \pm \hat \Delta_i$. 
By rotating the defect spins individually on a Bloch sphere with polar angles $\theta_i$ and azimuthal angles $\phi_i$ it is straightforward to demonstrate that the defect Hamiltonians become

\begin{equation}
\hat H_{\text{defects}}^{(i)} = \bar H_i + \hat \Delta_i \otimes R_i, \quad R_i = \begin{bmatrix}
\cos\theta_i & e^{-i\phi_i}\sin\theta_i\\
e^{i\phi_i}\sin\theta_i & -\cos\theta_i
\end{bmatrix}\,,
\label{eq:defect_hamiltonian}
\end{equation}

\noindent where $\bar H_i = (\hat H_{\text{defect},\uparrow}^{(i)} + \hat H_{\text{defect},\downarrow}^{(i)})/2$ is the mean Hamiltonian of the spin channels and $2\hat \Delta_i= \hat H_{\text{defect},\uparrow}^{(i)} - \hat H_{\text{defect},\downarrow}^{(i)}$ is the difference.
 This way of rotating the defect spins is equivalent to the method used in Ref.~\cite{wilhelm2015ab}, except that this explicitly allows us to rotate each defect spin independently of the others. It follows from Eq.~\ref{eq:defect_hamiltonian} that spin flipping only occurs when the electron spin is not aligned with the defect spin, and the coupling between spin-channels is proportional to $\hat\Delta_i$. The spin-dependent transport is thus equivalent to having two separate channels that couple when $\hat \Delta$ is non-zero, such as at magnetic impurity sites as illustrated in Fig.~\ref{fig:spin_transport_device}.

The transmittance between any two leads $p$ and $q$ of a multi-terminal system can be calculated using the Landauer-Büttiker formula~\cite{datta1995electronic} $T_{pq} = \text{Tr}\{\Gamma_pG\Gamma_qG^\dagger\}$, where $G=[(E+i\varepsilon)I- H-\sum_n\Sigma_n]^{-1}$ is the retarded Green's function. $\Sigma_n$ and $\Gamma_n$ are the self-energy and linewidth functions, respectively, of lead $n$. A small imaginary part $\varepsilon=10^{-5}$~eV is added to the energy for numerical stability. If the spins are decoupled in the leads, it is easy to demonstrate that the spin-channel resolved transmittance between the leads of a spin-dependent two-terminal system becomes~\cite{pareek2002spin,wilhelm2015ab}

\begin{equation}
T_{\sigma,\sigma'} = \text{Tr}\{\Gamma^{(L)}_\sigma G\Gamma^{(R)}_{\sigma'} G^\dagger\}\,,
\end{equation}

where $\Gamma^{(L)}_\sigma$ ($\Gamma^{(R)}_{\sigma'}$) is the linewidth function of the left (right) lead with spin $\sigma$ ($\sigma'$). The transmittance on this form can be computed efficiently using the recursive Green's functions (RGF) technique (as outlined in Refs.~\cite{markussen2006electronic,mackinnon1985calculation}). All calculations are performed on unit cells with a relatively large width of 12.8 nm in order to minimize finite size effects. Furthermore, the calculations are performed using periodic boundary conditions transverse to the transport direction and the results are averaged over 29 $k$-points. Exactly at the CNP, the only propagating mode in the leads is at $k=0$. It is therefore important to ensure that this is included.

The spin-conserved transport is defined as $T_{sc} = T_{\uparrow\uparrow}+T_{\downarrow\downarrow}$ and the spin-flipped transport is defined as $T_{sf} = T_{\uparrow\downarrow}+T_{\downarrow\uparrow}$. We expect the total transport $T=T_{sc}+T_{sf}$ to be either Ohmic or localized. For Ohmic transport the resistance is $R(L) = R_c + R_s L/W$, where $R_c$ is the contact resistance, $R_s$ is the sheet resistance, $L$ is the device length and $W$ is the width of the unit cell. In the localization regime, the resistance is $R(L) = R_c \exp(L/\xi)$, where $\xi$ is the localization length. By fitting the total transport to a compound expression
\begin{equation}
R(L) =  \frac{h}{2e^2T} = R_c\exp(L/\xi)+R_s L/W\,,
\label{eq:resistance}
\end{equation}

\noindent we obtain both localization length and Ohmic resistance. In the limits $\xi\rightarrow \infty$ and $R_s\rightarrow 0$, this expression reduces to the Ohmic and localization regimes, respectively.

We can use the spin polarization $P$ to obtain the spin relaxation length $\lambda_S$. According to Zurek \emph{et al.} \cite{zurek2007gaussian}, the spin relaxation mechanism can be either exponential or Gaussian, depending on the distribution of spin-couplings to an environment. In order to include both cases as well as any intermediate relaxation mechanism, we fit the spin polarization according to the following expression

\begin{equation}
P(L) = \frac{T_{sc}(L)-T_{sf}(L)}{T_{sc}(L)+T_{sf}(L)} = e^{-(L/\lambda_S)^n}\,.
\label{eq:spin_polarization}
\end{equation}

\noindent It follows that the spin relaxation behavior is exponential when $n=1$ and Gaussian when $n=2$.

\section{Results}
\begin{figure}[h]
	\centering
	\includegraphics[width=0.95\columnwidth]{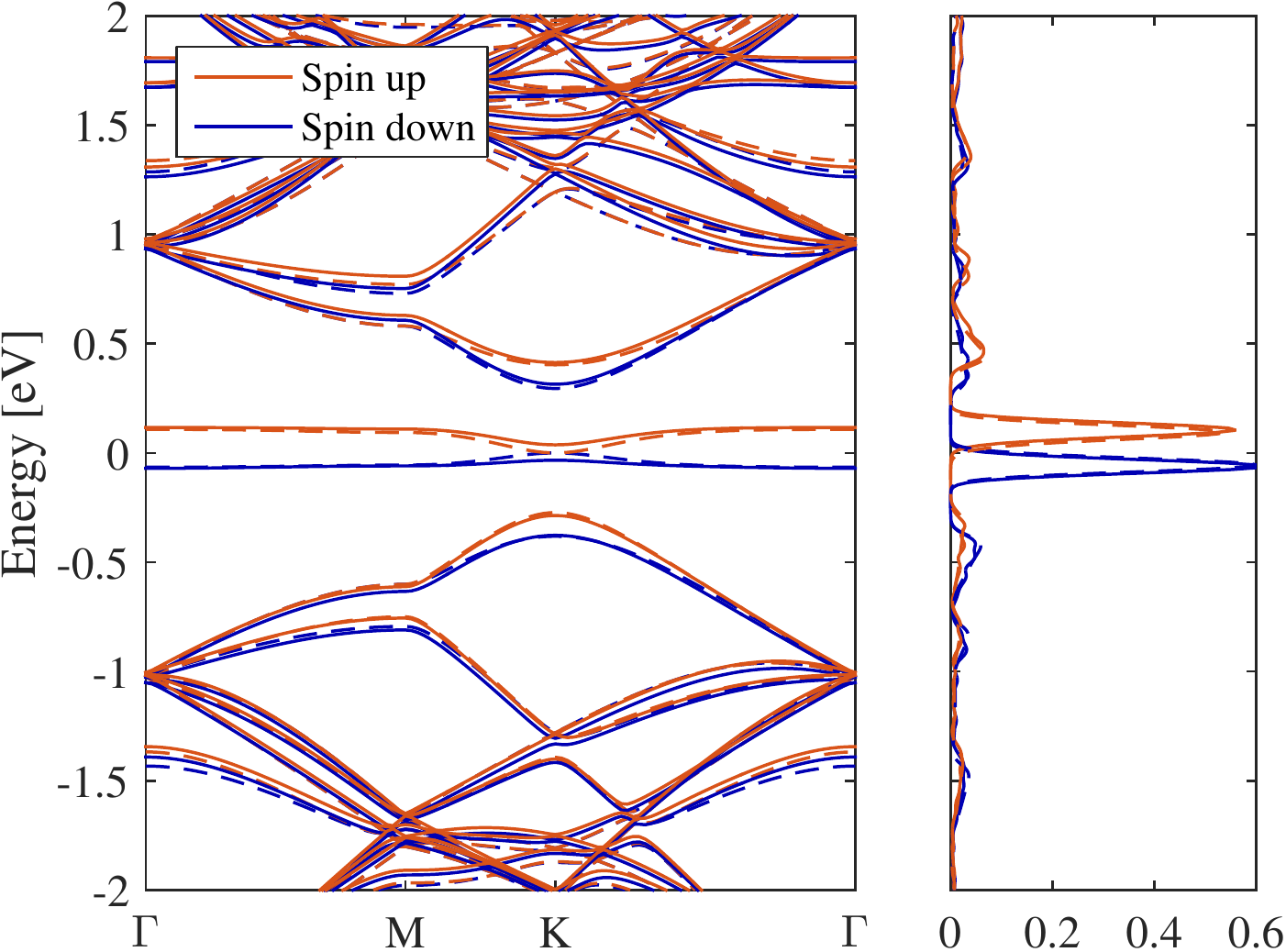}
	\caption{Spin-dependent band structure (left) and local density of states on the H atom (right) for an $8\times 8$ graphene unit cell with one H adatom. Solid lines are DFT results and dashed lines are TB results.}
	\label{fig:tbspintransport_H_adatom_bandstructure}
\end{figure}

By fitting to the pristine graphene DFT band structure, we find the C-C hopping parameters $t_1 = -2.855$ eV, $t_2 = -0.185$ eV and $t_3=-0.190$ eV for the first, second and third nearest neighbors, respectively. The $t_1$ and $t_2$ parameters are fitted freely, and $t_3$ is included by assuming
that $t_3 = t_1 (0.18/2.7)$, where the factor is motivated by earlier models \cite{hancock2010generalized}. The C on-site energy is vanishing. By fitting to the band structure of a system with an H adatom, the nearest-neighbor C-H hopping parameter can be taken as spin-independent with a value of 9.475 eV. The on-site energy at the hydrogen site is spin-dependent with a value of 4.689 eV for the majority spin component and 1.853 eV for the minority spin component. As the only spin-dependent parameter is on H adatoms, spin-flipping only occur at these sites in our model according to Eq.~\ref{eq:defect_hamiltonian}. The fitted band structure and LDOS are shown in Fig.~\ref{fig:tbspintransport_H_adatom_bandstructure}, where they are compared to the DFT calculations. The figure shows excellent agreement between TB and DFT band structures as well as the TB LDOS and the DFT pDOS.

\begin{figure}
\includegraphics[width=0.95\columnwidth]{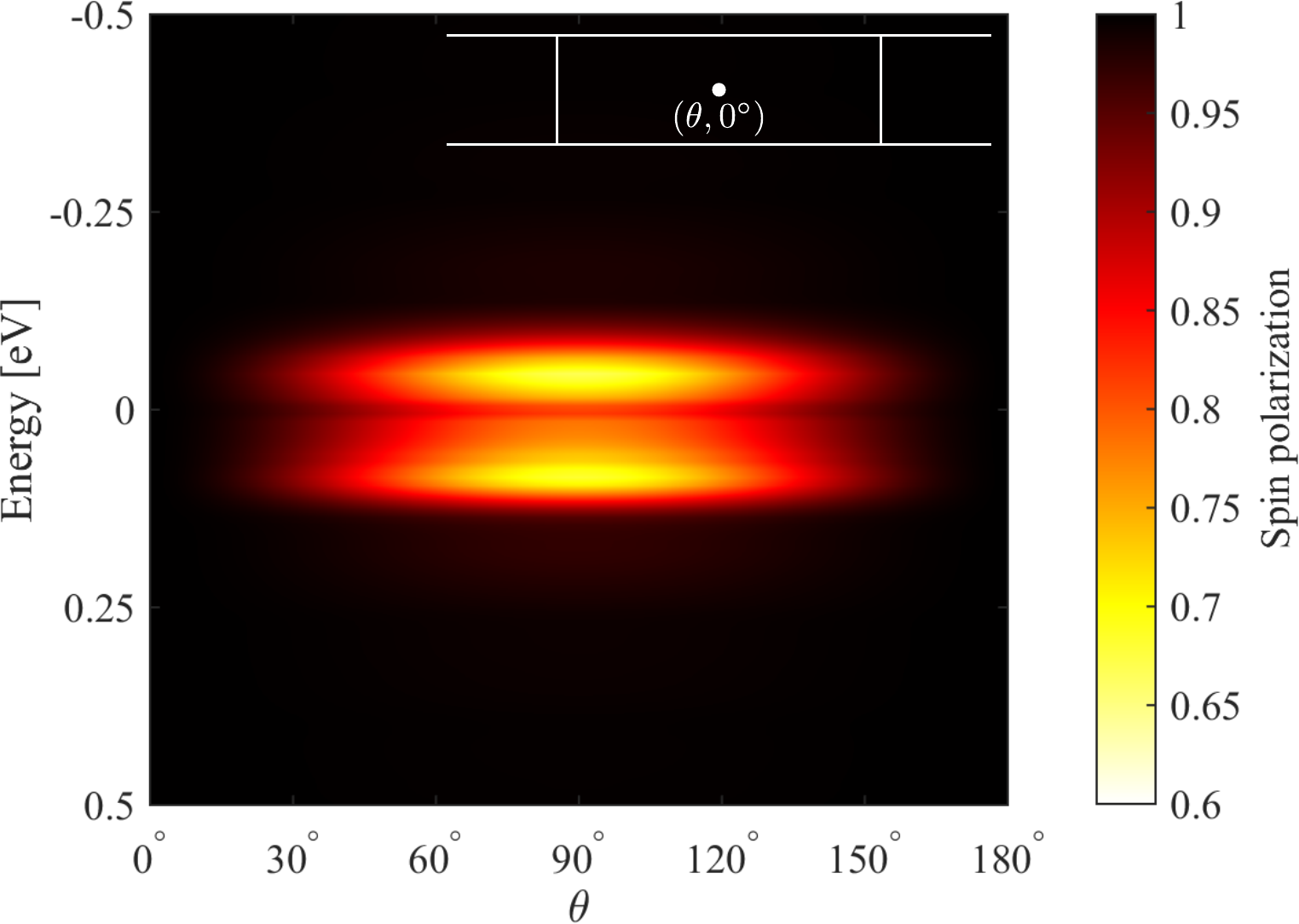} 
\caption{Spin polarization as a function of energy and angle of a graphene system with a single H adatom. The inset shows an illustration of the device.} 
\label{fig:Single_H_adatom_polarization} 
\end{figure}

The spin polarization of a system containing a single H adatom is shown in Fig.~\ref{fig:Single_H_adatom_polarization}. When there is only a single defect, the transport properties do not depend on the azimuthal defect spin angle $\phi$. Therefore, only the polar angle $\theta$ and the energy $E$ are varied. The figure shows that the spin scatters very strongly near the charge neutrality point (CNP), $E\simeq 0$, resulting in a significantly decreased spin polarization. This is a consequence of scattering on H adatoms, which as defect bands that span approximately $\pm 0.1$ eV around the CNP, cf.~Fig.~\ref{fig:tbspintransport_H_adatom_bandstructure}. This means that a single H adatom with in-plane defect spin is able to destroy almost half of the spin polarization for energies near the CNP. This is in good agreement with Wilhelm \emph{et al.}~\cite{wilhelm2015ab}, who found that an $N=11$ armchair graphene nanoribbon with a single H adatom with in-plane spin can have spin-flip transmittance that can surpass the spin-conserved part.

\begin{figure}
\includegraphics[width=0.95\columnwidth]{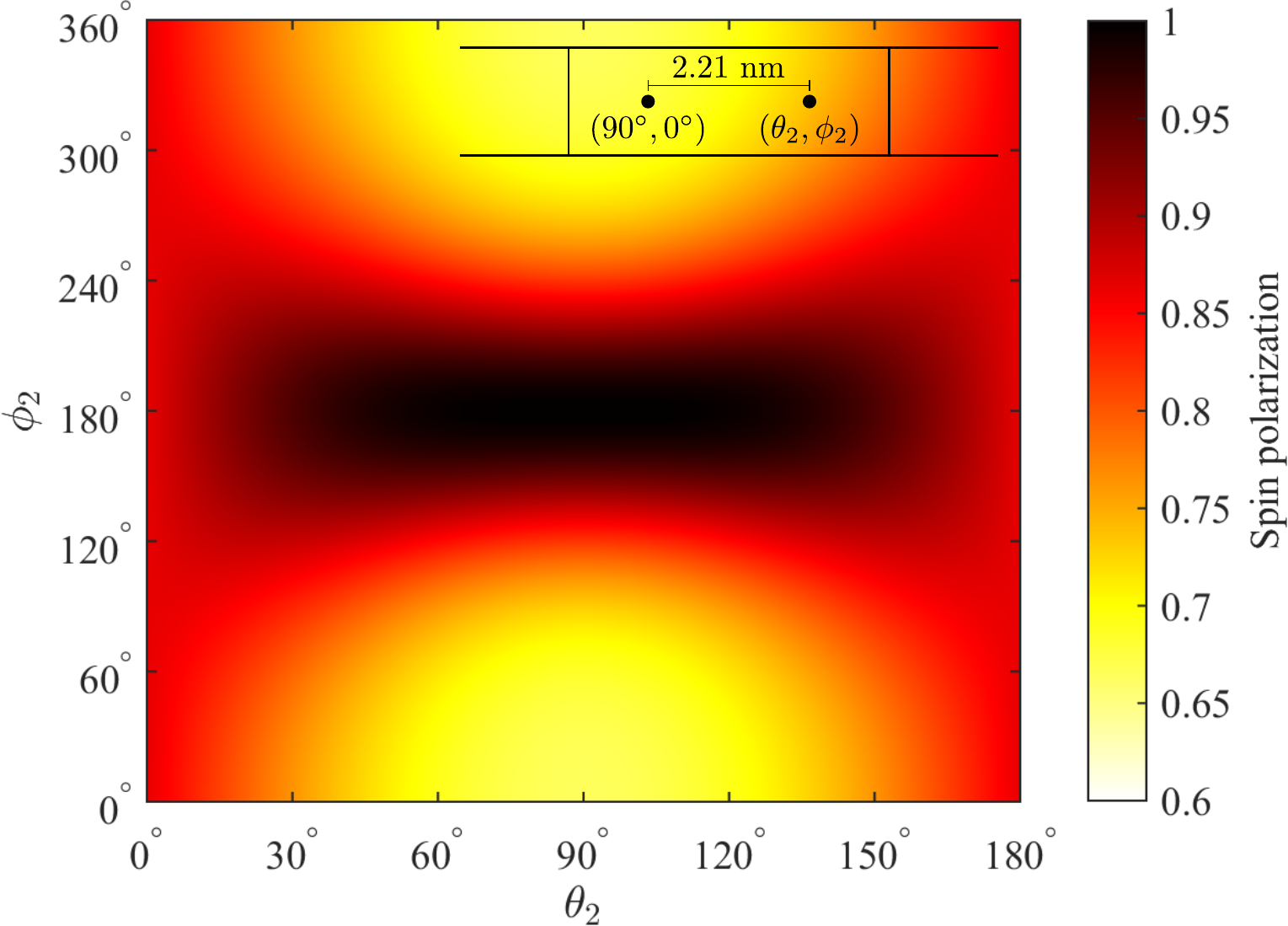} 
\caption{Spin polarization of a system with two H adatoms with defect spin angles $(\theta_1,\phi_1) =(90^\circ,0^\circ)$ and $(\theta_2,\phi_2)$, respectively, at an energy of $E=0.0$ eV. The H adatoms are placed on a line parallel to the transport direction 2.21 nm apart. The inset shows an illustration of the device.} 
\label{fig:Two_H_adatoms_polarization} 
\end{figure}

In order to obtain information on the interference effects of spin-flipping, we calculate the spin polarization of a system with two H adatoms separated by a distance of 2.21 nm parallel to the transport direction, see Fig.~\ref{fig:Two_H_adatoms_polarization}. The spin polarization is evaluated at the CNP and the orientation of the defect spins have been chosen to be $(\theta_1,\phi_1) = (90^\circ,0^\circ)$ and $(\theta_2,\phi_2)$, respectively. The figure shows that the total spin polarization is minimal when the defect spins are in-plane and point in the same direction, whereas it is maximal, when the spins are in-plane and point in opposite directions. 
In Fig.~\ref{fig:Single_H_adatom_polarization}, we saw that a single defect with in-plane spin could flip almost half of the electron spin to the opposite channel. Now we see that by having two defects with oppositely oriented in-plane spins, the second can almost completely negate the first spin flip. 
When the two defect spins point in opposite directions, the phase change associated with spin-flipping will have equal size and opposite sign. This means that the electron spin will be in phase with the input spin after the second spin-flip, leading to constructive interference. 
This is not the case when the defect spins point  in the same direction. The interference between defects is thus very important and should not be ignored.

\begin{figure}
\includegraphics[width=1\columnwidth]{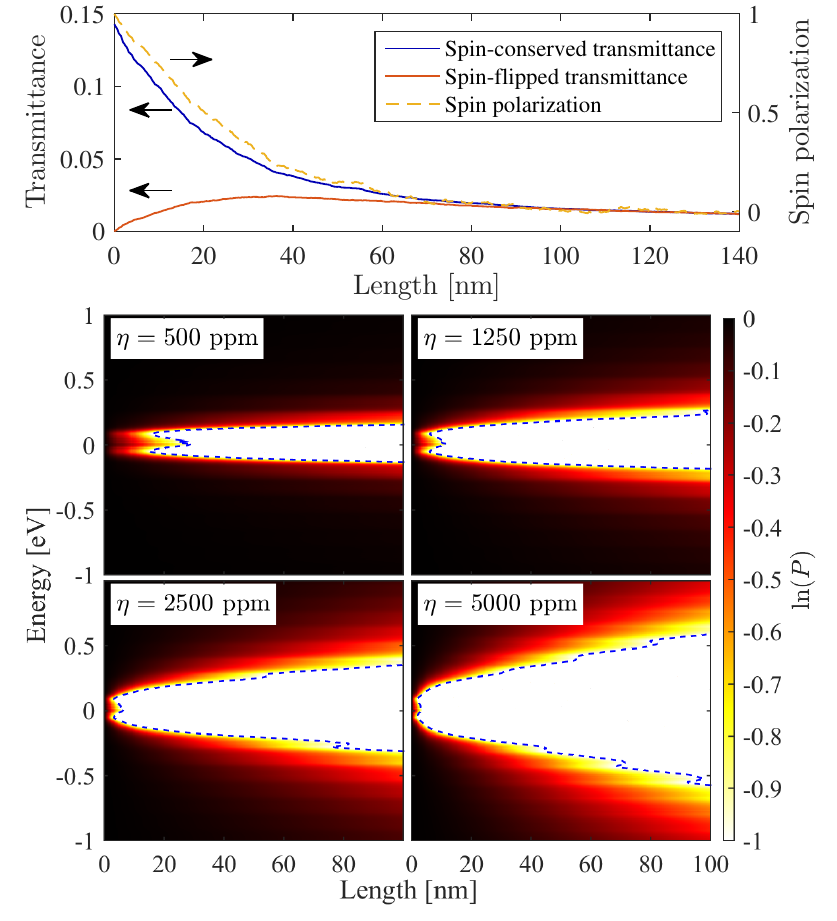} 
\caption{Ensemble-averaged transmittance and spin polarization as a function of length (top) for a system with impurity concentration $\eta = 500$ ppm calculated at the CNP. Ensemble-averaged spin polarization as a function of energy and length (bottom) for different impurity concentrations. The dashed lines show the spin relaxation length.} 
\label{fig:spinpolarization} 
\end{figure}

\begin{figure}
\includegraphics[width=0.95\columnwidth]{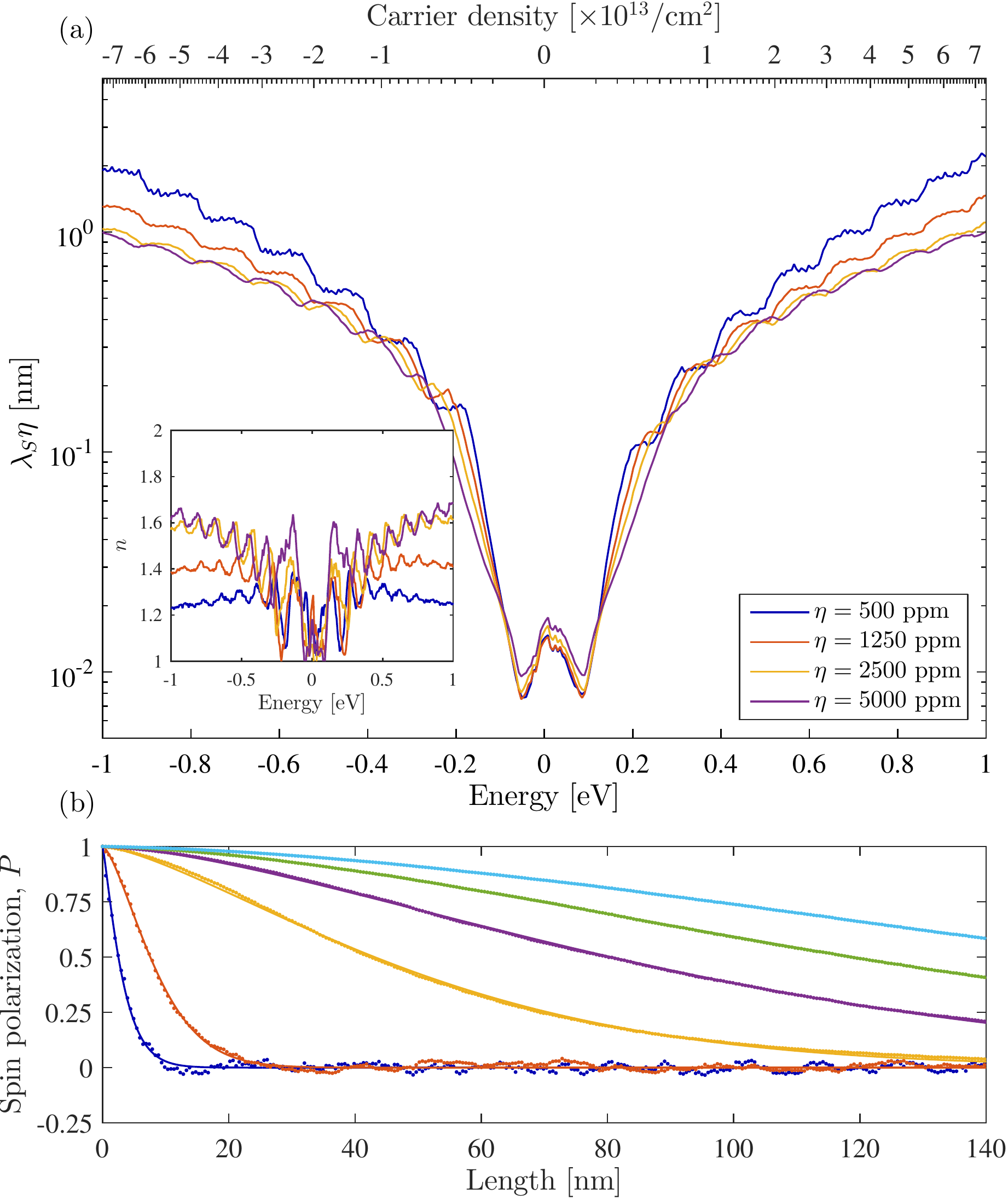} \\
\caption{(a) Normalized spin relaxation length $\lambda_S\eta$ and exponent $n$ (inset) obtained by fitting against Eq.~\ref{eq:spin_polarization}. The spin relaxation lengths are normalized with the defect concentration in order to illustrate their nearly linear scaling with respect to it.  (b) Examples of fitting the spin relaxation against Eq.~\ref{eq:spin_polarization} of a system with $5000$~ppm H adatoms for different energies. The energies are between 0.0~eV (fastest decay) and 1.0~eV (slowest decay) in steps of 0.2~eV. The dots are the ensemble averaged spin polarizations and the lines are the corresponding fitted functions. } 
\label{fig:H_adatom_spinrelaxation} 
\end{figure}

We now turn to calculating the effects of multiple magnetic hydrogen adatoms on graphene. We place hydrogen adatoms at random positions uniformly distributed across the device according to a predefined impurity concentration $\eta$. The impurity concentration is related to the impurity density by $n=4\eta/(\sqrt{3}a^2)$, where $a$ is the graphene lattice constant. 
We wish to keep the device non-magnetic in order to isolate spin relaxation from other magnetic effects. 
Therefore, we choose the directions of the defect spins at random, uniformly distributed on a Bloch sphere.
The transport is calculated for very long devices of 147.5~nm, which contain a total of 72,000 carbon atoms in the unit cell. Using the RGF method, we can extract the transport after each slice of the device, allowing us to obtain the length-dependent transport directly. 
In order to minimize the effects of the finite width of the unit cell, we average over an ensemble of 150 device realizations. The spin polarization as a function of length and energy for different impurity concentrations is shown in Fig.~\ref{fig:spinpolarization} as well as an example of transmittance and spin polarization as a function of length for a single energy and impurity concentration. We show the logarithm of the spin polarization in the range between -1 and 0 in order to highlight the spin relaxation length, which is defined as the length at which $\ln(P(L))=-1$. The figure shows that the spin polarization decays very fast for energies close to the H adatom defect bands, cf.~Fig.~\ref{fig:tbspintransport_H_adatom_bandstructure}. As expected, the spin polarization decays faster with increasing impurity concentration. Note that the spin polarization also decays for energies away from the H defect bands, due to the relatively small spin splitting in the remaining band structure. The small energy-dependent oscillations in Figs.~\ref{fig:spinpolarization} and \ref{fig:H_adatom_spinrelaxation} are due to finite size effects originating from the finite width of the unit cell.

By fitting the spin polarization to Eq.~\ref{eq:spin_polarization} we obtain the spin relaxation length as well as the exponent $n$, which provides information on the spin relaxation mechanism, see Fig.~\ref{fig:H_adatom_spinrelaxation}a. A few examples of the fitting procedure is included in Fig.~\ref{fig:H_adatom_spinrelaxation}b in order to illustrate the excellent quality of the fits. The carrier concentration on the figure is computed at Fermi energies corresponding to the energy axis. Positive carrier densities refer to electron doping and negative carrier densities refer to hole doping. The spin relaxation length is very short for energies near the H defect bands. For the same energies, the spin relaxation mechanism is predominantly exponential with an exponent of $n\simeq1$. For energies further away from the CNP, the spin relaxation length increases. We note that $\lambda_S$ has two minima near the CNP, which are correlated with the large spin-splitting of the H adatom defect bands. Exactly at the CNP, the spin-splitting of the defect bands is vanishing, resulting in a local maxima. 
The figure shows that there is an almost linear scaling of the spin relaxation length with respect to impurity concentration, especially near the CNP. Away from the CNP we observe that $n$ decreases with decreasing impurity concentration. This suggests that the spin relaxation mechanism tends toward exponential (Markovian) behavior in the highly dilute impurity limit. 
Importantly, we see that the decay of the spin polarization as a function of length need not be exponential nor Gaussian, which means that any good theory on spin relaxation should not presume anything about the spin relaxation behavior, except in the limit of very dilute systems, where the approximation of exponential decay seems to be valid. For energies near the CNP, the normalized localization length is $\lambda_S\eta \approx 0.01$ nm. In order to obtain experimentally observed spin relaxation length of about $\lambda_S \simeq 2$ $\mu$m~\cite{wojtaszek2013enhancement}, the impurity concentration should be $\eta \approx 5$~ppm, which is more than an order of magnitude larger than the prediction by Kochan \emph{et al.}~\cite{kochan2014spin} of 0.36 ppm. 
We expect our model to be more accurate as it is based on a full transport calculation and therefore takes interference effects into account.
Our prediction of the impurity concentration is, however, in closer agreement with Soriano \emph{et al.}~\cite{soriano2015spin}, who found that an impurity concentration of $15$~ppm gives spin relaxation times in agreement with experiment based on time propagation of the spin polarization operator using a self-consistent Hubbard model. 

\begin{figure}
\includegraphics[width=0.95\columnwidth]{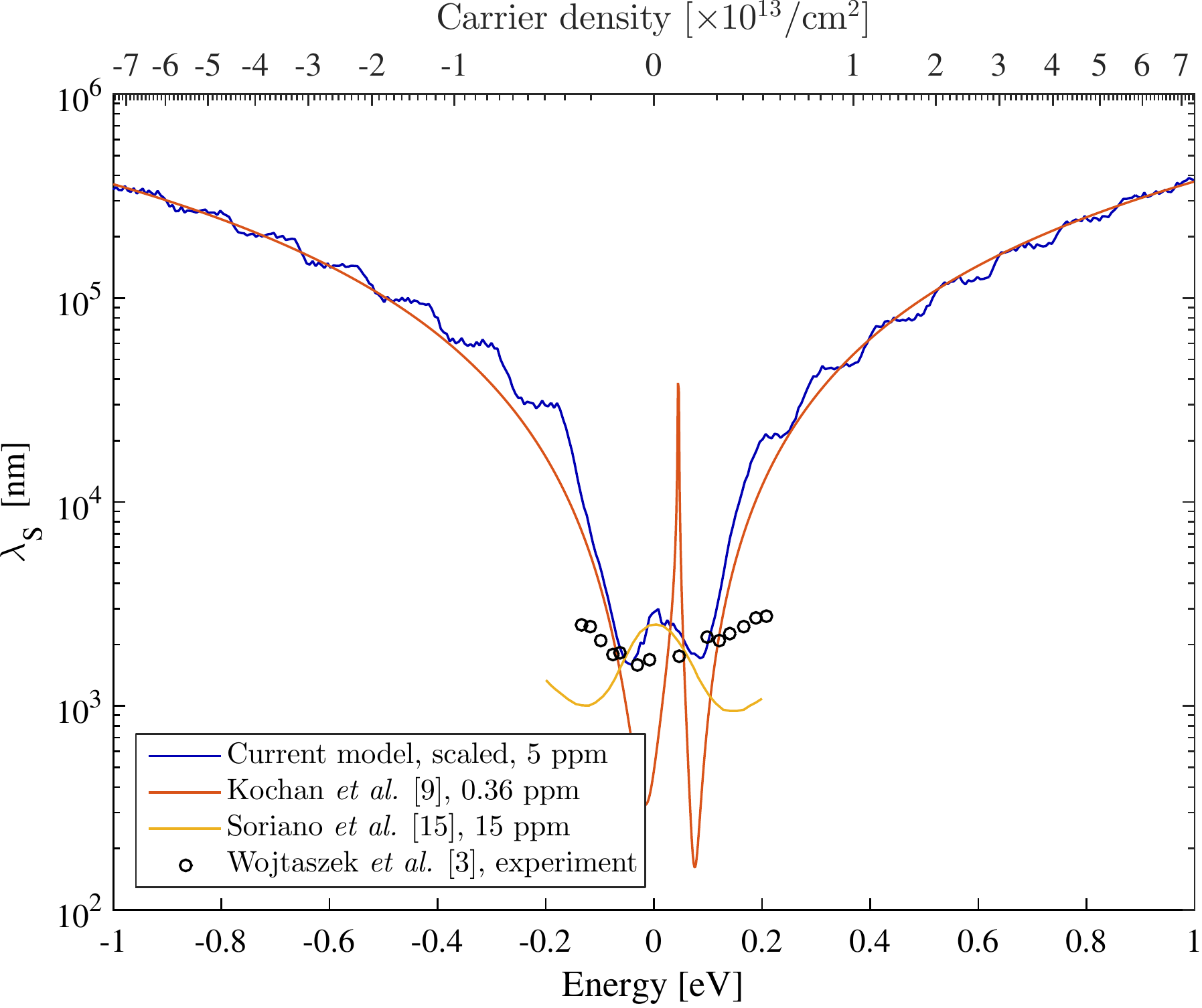} \\
\caption{Comparison of spin relaxation lengths obtained by different authors.} 
\label{fig:H_adatom_spinrelaxation_comparison} 
\end{figure}

A comparison of spin relaxation lengths obtained by the current model and those obtained by other theoretical methods and experiment is presented in Fig.~\ref{fig:H_adatom_spinrelaxation_comparison}. We have rescaled our 500 ppm result to 5 ppm by multiplying it by a factor of 100. The two other theoretical results \cite{kochan2014spin,soriano2011magnetoresistance} have calculated the spin relaxation time $\tau_S$, which is related to the spin relaxation length by $\lambda_S=v_S\tau_S$ in the ballistic regime and by $\lambda_S = \sqrt{D_S\tau_S}$ in the diffusive regime, where $v_S$ is spin carrier velocity and and $D_S$ is the spin diffusion constant. In the low-defect-density case, we expect to be in the ballistic regime. Therefore we compare results that are all obtained in the low-defect-density case. We obtain a velocity $v_S=1.65\times10^4$~m/s by least-of-squares fitting between our result and the analytic result obtained by Kochan \emph{et al.} We observe that the result by Kochan \emph{et al.} is in fairly good agreement ours regarding the location of the two minima near the CNP and in quantitative agreement further away from the CNP. However, their result predicts variations over several orders of magnitude near the CNP, whereas our result predicts a variation of only about a factor of 2. In fact, their result is singular exactly at the CNP. Furthermore, we compare with experimental results of hydrogenated graphene obtained by Wojtaszek \emph{et al.} \cite{wojtaszek2013enhancement}. Note that the experimental results were obtained without detailed knowledge of the defect concentration. However, the authors estimated the concentration to be $200$~ppm. Lastly, we compare our results to the theoretical result by Soriano \emph{et al.} \cite{soriano2011magnetoresistance}. The figure shows that their result is neither in qualitative agreement with our model nor the analytic result by Kochan \emph{et al.} or experiment. We speculate that the deviation arises from the fact that Soriano \emph{et al.} uses vacancies in graphene to model hydrogen adatoms, whereas both our model and the model used by Kochan \emph{et al.} employ a parametrization of hydrogen on graphene.
Theoretical predictions \cite{soriano2015spin,kochan2014spin,tuan2014pseudospin} including our own, show that the spin relaxation time (or spin relaxation length) decreases with increasing impurity concentration. However, experimental work on hydrogenated graphene shows that the spin relaxation time (or spin relaxation length) actually increases with increasing impurity concentration~\cite{wojtaszek2013enhancement}. The origin of this discrepancy remains elusive, but could stem from interaction between graphene and the substrate, as this has not been included in any of the theoretical models.

\begin{figure}
\includegraphics[width=0.95\columnwidth]{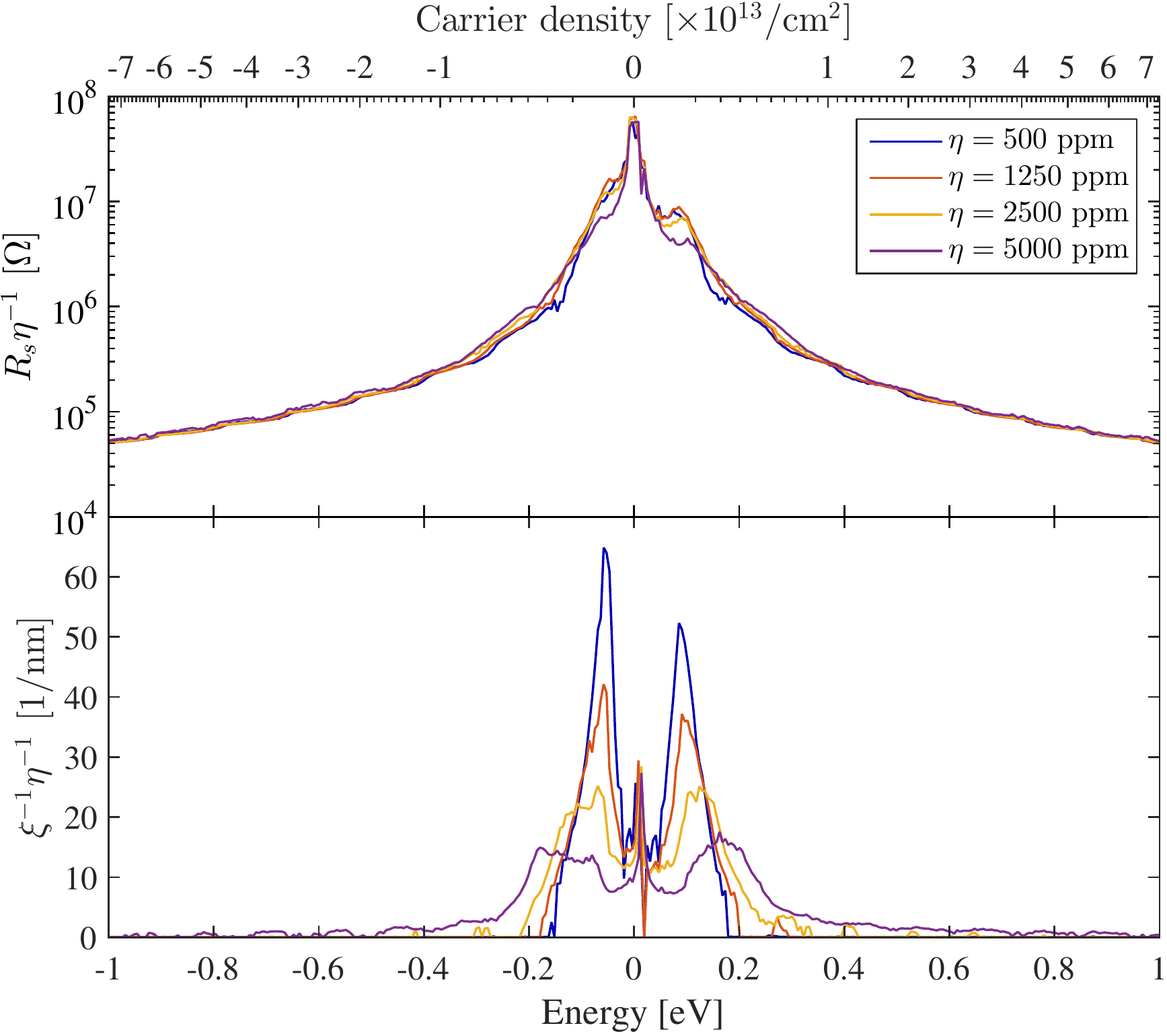} 
\caption{Normalized Ohmic sheet resistance $R_s/\eta$ (top) and normalized localization length $\xi/\eta$ (bottom) obtained by fitting against Eq.~\ref{eq:resistance}} 
\label{fig:H_adatom_totaltransport} 
\end{figure}

By fitting the total transmittance against Eq.~\ref{eq:resistance} we obtain the Ohmic sheet resistance as well as the localization length, see Fig.~\ref{fig:H_adatom_totaltransport}. We observe localization near the H defect bands, cf.~Fig.~\ref{fig:tbspintransport_H_adatom_bandstructure}, and vanishing localization elsewhere. Additionally, the figure shows that the sheet resistance scales linearly with respect to impurity concentration. However, the scaling of the localization length is far from linear, which shows that the localization induced per atom decreases with increasing impurity concentration. Furthermore, as the impurity concentration is decreased the energy window, at which there is localization narrows.

\section{Conclusions}
We have demonstrated that the Landauer-Büttiker formalism can be used to calculate spin-dependent transport of systems with magnetic impurities with individually oriented magnetic moments. In this work, we study hydrogen adatoms on graphene. By calculating the length-dependent transport, we can extract properties such as spin relaxation length, localization length and sheet resistance. We have shown that there is strong localization for energies around the hydrogen-induced defect bands, which also leads to a very high sheet resistance. Away from the defect bands there is vanishing localization. Furthermore, we have demonstrated that the spin relaxation length is very short for energies around the hydrogen-induced defect bands and that the spin relaxation mechanism is exponential (Markovian) near the CNP and non-exponential (non-Markovian) otherwise. Additionally, we have shown that spin relaxation length and sheet resistance scale nearly linearly with impurity concentration, whereas the localization length does not.

\section*{Acknowledgments}
M. R. Thomsen and T. G. Pedersen gratefully acknowledge the financial support from the Center for Nanostructured Graphene (Project No. DNRF58) financed by the Danish National Research Foundation and from the QUSCOPE project financed by the Villum Foundation. 
The work by A. Harju and M. Ervasti has been supported by the Academy of Finland through its Centres of Excellence Program (project no. 251748). We acknowledge the computational resources provided by Aalto Science-IT project and Finland's IT Center for Science (CSC).

\bibliography{literature}

\end{document}